\documentclass[9pt,twocolumn,twoside]{opticajnl}
\journal{opticajournal} 

\setboolean{shortarticle}{true}

\usepackage{lineno}

\title{Si-waveguide-based optical power monitoring of 2$\times$2 Mach--Zehnder interferometer based on InGaAsP/Si hybrid MOS optical phase shifter}

\author[1,*]{Tomohiro Akazawa}
\author[2]{Stéphane Monfray}
\author[2]{Frédéric Boeuf}
\author[1]{Kasidit Toprasertpong}
\author[1]{Shinichi Takagi}
\author[1]{Mitsuru Takenaka}

\affil[1]{Department of Electrical Engineering and Information Systems, The University of Tokyo, Tokyo 113-8656, Japan}
\affil[2]{STMicroelectronics, 38920 Crolles, France}

\affil[*]{akazawa@mosfet.t.u-tokyo.ac.jp}

\begin{abstract}
Transparent in-line optical power monitoring in Si programmable photonic integrated circuits (PICs) is indispensable for calibrating integrated optical devices such as optical switches and resonators. A Si waveguide (WG) photodetector (PD) based on defect-mediated photodetection is a promising candidate for a transparent in-line optical power monitor owing to its simplicity and ease of integration with a fully complementary metal-oxide-semiconductor (CMOS)-compatible process. Here, we propose a simple optical power monitoring scheme for a 2$\times$2 Mach--Zehnder interferometer (MZI) optical switch based on InGaAsP/Si hybrid MOS optical phase shifters. In the proposed scheme, a low-doped p-type Si WG PD with a response time of microseconds is utilized as a transparent in-line optical power monitor and the ground terminal of the MOS optical phase shifter is shared with that of the Si WG PD to enable the simple monitoring of the output optical power of the MZI. Based on this scheme, we experimentally demonstrate that the output optical power of a 2$\times$2 MZI can be simply monitored by applying a bias voltage to the Si slabs formed at the output WGs of the MZI without excess optical insertion loss.
\end{abstract}

\setboolean{displaycopyright}{false} 

\begin{document}
\maketitle

\noindent\textbf{Introduction.} 
With the rapid growth of the Si photonics platform, Si programmable photonic integrated circuits (PICs) have been extensively developed for many promising applications such as communication \cite{Suzuki2019, Sun2015}, computing \cite{Shen2017,Qiang2018}, and sensing \cite{Zhang2022}. Those PICs often employ numerous 2$\times$2 Mach--Zehnder interferometers (MZIs) with optical phase shifters and 3 dB couplers, and the reconfigurability of a PIC is achieved by electrically controlling the optical phase of the propagating light in a Si waveguide (WG). Therefore, the accurate control of phase shift values at each phase shifter is crucial for reconfiguring a PIC for a specific purpose \cite{Bogaerts2020}. To achieve that in a PIC where numerous MZIs are cascaded, the output optical power at each MZI needs to be monitored through in-line optical power monitors (OPMs) for the feedback control of the MZIs. As such an in-line OPM, a low-loss transparent photodetector (PD) with a simple device structure is desired \cite{Bogaerts2020}. A common method to achieve an in-line OPM is to employ a germanium (Ge) PD with optical power tapping \cite{Pantouvaki2013}. Although this method can be easily integrated into a PIC, it suffers from a relatively large optical insertion loss caused by optical power tapping. 
A promising approach to achieving a transparent in-line OPM is to utilize a Si WG PD, in which the Si WG core itself is used for photodetection, offering many advantages such as design simplicity, ease of integration with a fully complementary metal-oxide-semiconductor (CMOS)-compatible process. Although Si does not absorb infrared light with a wavelength above 1100 nm, defect states or surface states in Si can be utilized for the detection of the infrared light. Recently, a transparent in-line OPM called CLIPP, which utilizes the conductance change in a Si WG induced by surface state absorption, has been demonstrated \cite{Morichetti2014,Grillanda2014}. CLIPP does not require any metal contacts with the Si WG layer, which is advantageous for applications where metal contacts for the Si WG layer are unnecessary such as the PICs based on thermo-optic phase shifters \cite{Grillanda2014}. However, one drawback is that the small change in WG conductance needs to be detected by a complex lock-in amplifier through capacitive coupling. Therefore, in the case of the PICs where contacts for the Si layer are required, using Si WG PDs with contacts for the Si layer is more beneficial for simplifying the reading of optical power.

\begin{figure}[H]
\centering
\includegraphics[width=\linewidth]{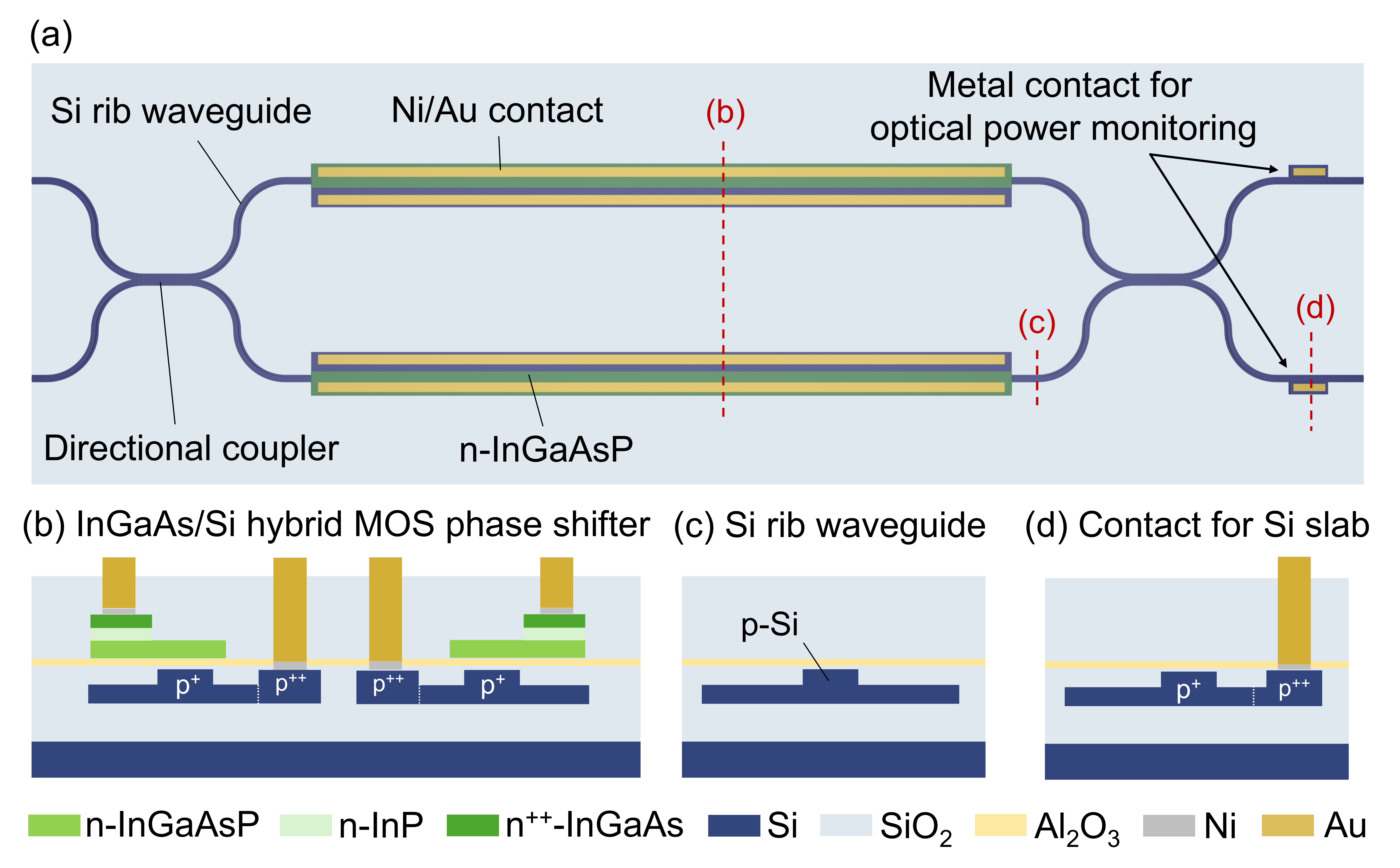}
\caption{(a) Plan-view schematic of the proposed simple transparent in-line power monitoring scheme using Si WG PDs. The ground (GND) terminals of InGaAsP/Si hybrid MOS optical phase shifters work also as the GND terminals for Si WG PDs, enabling optical power monitoring by simply applying bias voltage to the Si slabs at the output ports of the 2$\times$2 MZI without excess optical insertion loss. Cross-sectional schematics of (b) InGaAsP/Si hybrid MOS optical phase shifter, (c) low-doped p-Si rib WG, and (d) Si rib WG with $\mathrm{p}^{++}$-Si slab for metal contact.}
\label{fig:fig1}
\end{figure}

Here, we propose a simple transparent in-line power monitoring scheme based on Si WG PDs with contacts for the Si layer, as shown in Fig. \ref{fig:fig1}(a), and demonstrate the power monitoring of a 2$\times$2 MZI with InGaAsP/Si hybrid MOS optical phase shifters \cite{Han2017}, as shown in Fig. \ref{fig:fig1}(b). Common defect-mediated Si WG PDs utilize ion implantation to introduce many defect states in Si to enhance the responsivity \cite{Bradley2005,Liu2006,Geis2007,Yu2012}. However, this results in excess optical insertion loss, making them unsuitable for transparent optical power monitoring. In this work, we used a low-doped p-type Si WG as a PD, as shown in Fig. \ref{fig:fig1}(c), since light can still be detected owing to inherently induced defect states in Si \cite{Bradley2005,Perino2022}, enabling fully transparent in-line optical power monitoring. Moreover, to simplify the optical power monitoring scheme in a 2$\times$2 MZI, the ground (GND) terminals on the Si slab for the InGaAsP/Si hybrid MOS optical phase shifter were shared with those for the Si WG PDs, and by simply applying a bias voltage on the Si slabs at the output ports of the 2$\times$2 MZI as shown in Fig. \ref{fig:fig1}(d), we demonstrated that the output light power of the 2$\times$2 MZI can be monitored in a simple manner without excess optical insertion loss.
\textbf{Design and fabrication.} In this work, a Si rib WG with a 450-nm-wide and 300-nm-thick core, and 1.3-µm-wide and 160-nm-thick slabs was used. For metal contacts to Si WG PDs and optical phase shifters, 5.5-µm-wide and 5.35-µm-wide $\mathrm{p}^{++}$-doped slabs ($N_\mathrm{A} \approx 10^{19}\ \mathrm{cm}^{-3}$) were formed, respectively. Note that although the Si WG core in the power monitoring region in Fig. \ref{fig:fig1}(c) is p$^+$-doped ($N_\mathrm{A} \approx 10^{17}\ \mathrm{cm}^{-3}$) to lower resistance, the free carrier absorption associated with the doping is negligible owing to the small p$^+$-doped Si length of 20 µm. As for optical phase shifters in 2$\times$2 MZI, InGaAsP/Si hybrid MOS optical phase shifters were used since they enable efficient light modulation owing to the light effective electron mass of InGaAsP \cite{Han2017}. Furthermore, ultrathin 30-nm-thick InGaAsP membranes were employed to eliminate tapers while maintaining a high coupling efficiency, enabling a simple fabrication process \cite{Ohno2020,Ochiai2022,Akazawa2022,Akazawa2024}. Optical phase shifters were fabricated by the following process: First, III--V epitaxial layers including 30-nm-thick n-InGaAsP ($\lambda_\mathrm{g} = 1.3\ $µ$\mathrm{m},\ N_\mathrm{D} = 5\times 10^{15}\ \mathrm{cm}^{-3}$), 20-nm-thick n-InP ($N_\mathrm{D} = 1 \times 10^{16}\ \mathrm{cm}^{-3}$), and 100-nm-thick $\mathrm{n}^{++}$-InGaAs ($N_\mathrm{D} = 1 \times 10^{19}\ \mathrm{cm}^{-3}$) layers grown on an InP substrate were bonded via 8-nm-thick Al$_2$O$_3$ onto a Si-on-insulator (SOI) wafer with SiO$_2$-embedded Si rib WGs. After removing the InP substrate and etch-stop layers by wet etching, the III--V regions of $\mathrm{n}^{++}$-InGaAs and n-InP for contacts were defined by electron-beam (EB) lithography and selective wet etching. Here, the InP layer was inserted between the n$^{++}$-InGaAs and n-InGaAsP layers to etch these layers selectively. Subsequently, the InGaAsP regions for the phase shifters were defined by EB lithography and reactive ion etching, and the 615-nm-thick SiO$_2$ layer was then deposited for surface passivation by sputtering. Finally, contacts for the III--V and Si layers were formed by a Ni/Au metal stack by EB evaporation and the lift-off process.
\begin{figure}[H]
\centering
\includegraphics[width=70mm]{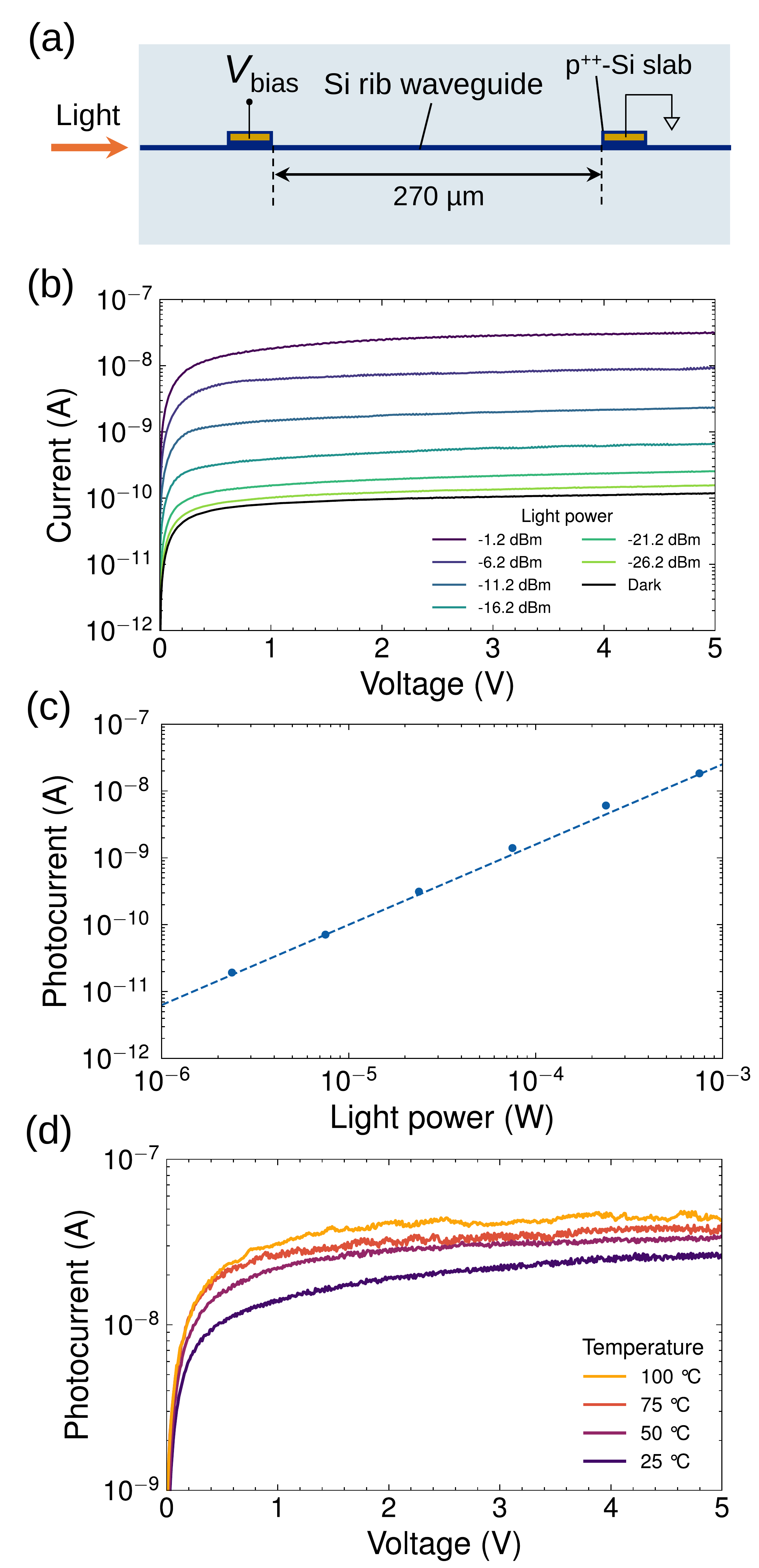}
\caption{(a) Plan-view schematic of a Si WG PD. The metal contacts were aligned parallel to the light propagation direction, and the distance between the metal contacts was 270 µm. (b) $I$--$V$ characteristics of the PD when $V_\mathrm{bias}$ = 1 V under different input light powers. (c) Photocurrent as a function of input light power. (d) Photocurrent of a Si WG PD as a function of bias voltage under different temperatures ranging from 25 $^\circ$C to 100 $^\circ$C.}
\label{fig:fig2}
\end{figure}
\textbf{Experimental results.} The photoresponse of the Si rib WG PD (Fig. \ref{fig:fig2}(a)) was characterized under continuous-wave (CW) light injection at a wavelength of 1550 nm from a tunable laser source (Santec, TSL-510). Here, the distance of the metal contacts for the PD was 270 µm. Note that the distance between the metal contacts and the Si WG core was $>$ 2 µm; therefore, the optical loss by metal absorption is negligible. The polarization of the input light was adjusted to the transverse electric (TE) mode of the Si WG by an in-line polarization controller, and the input light was coupled from a single-mode fiber to the Si rib WG via the grating coupler (GC) and then injected into a PD. Here, the input light power was controlled by an optical attenuator (Keysight, 81576A). By the cutback method, we estimated the propagation loss of a Si rib WG and the insertion loss of a GC to be 2.4 dB/cm and 3.2 dB, respectively. Fig. \ref{fig:fig2}(b) shows the $I$--$V$ characteristics of a PD measured by a semiconductor device analyzer (Agilent Technologies, B1500A). Note that the input power is defined here as the light power injected into the PD. Fig. \ref{fig:fig2}(c) shows the photocurrent as a function of the input light power. From this result, the responsivity was estimated to be about 2$\times$10$^{-5}$ A/W, a comparable value with previously reported unimplanted Si WG PDs \cite{Bradley2005,Perino2022}. From the responsivity value, assuming an internal quantum efficiency of 100\%, the absorption coefficient attributed to the photodetection was calculated to be 2.6$\times$10$^{-3}$ dB/cm at a wavelength of 1550 nm, a negligible value compared to a Si WG propagation loss of 2.4 dB/cm. Note that two photon absorption (TPA) is not responsible for the photodetection mechanism in our PDs as the photocurrent increases quadratically with the input light power in TPA, which was not observed in our results. Furthermore, we measured the $I$--$V$ characteristics of the PD at an input light power of -1.2 dBm under different temperatures ranging from 25 $^\circ$C to 100 $^\circ$C, and then the photocurrent was extracted as shown in Fig. \ref{fig:fig2}(d). Note that the photocurrent fluctuations at high temperatures can be attributed to the fluctuations of the coupling between the fiber and the GC due to thermal expansion. In defect-mediated PDs, carriers are excited by a two-step process: thermal and optical excitations \cite{Logan2009,Keevers1994}. The results in Fig. \ref{fig:fig2}(d) indicate that the thermal excitation rate increases with temperature, leading to the increase in photocurrent at high temperatures. Subsequently, the time response of the PD was characterized using the setup shown in Fig. \ref{fig:fig3}(a). A tunable laser was directly modulated by an electrical waveform generator (Agilent Technologies, 33522B), and pulse light signals at a wavelength of 1550 nm with a repetition frequency of 5 kHz were injected to the PD. Note that the rise and fall times of the input laser pulse were 624 ns and 626 ns, respectively. The electrical waveform from the PD was measured using a waveform generator/fast measurement unit (Agilent Technologies, B1530A) of the semiconductor device analyzer. Fig. \ref{fig:fig3}(b) shows the time response of the PD current under a bias voltage of 2 V when the input light power to the PD was 4.3 dBm. From Fig. \ref{fig:fig3}(b), we found that the rise and fall times of the PD were 2.1 µs and 3.4 µs, respectively, which were sufficiently short for the optical power monitoring application. Note that the rise and fall times here include those of the input laser pulse. We speculate that the response time of the PD is limited by the transit time of the photogenerated carriers due to a relatively large metal gap of 270 µm, as shown in Fig. \ref{fig:fig2}(a).
\begin{figure}[ht]
\centering
\includegraphics[width=80mm]{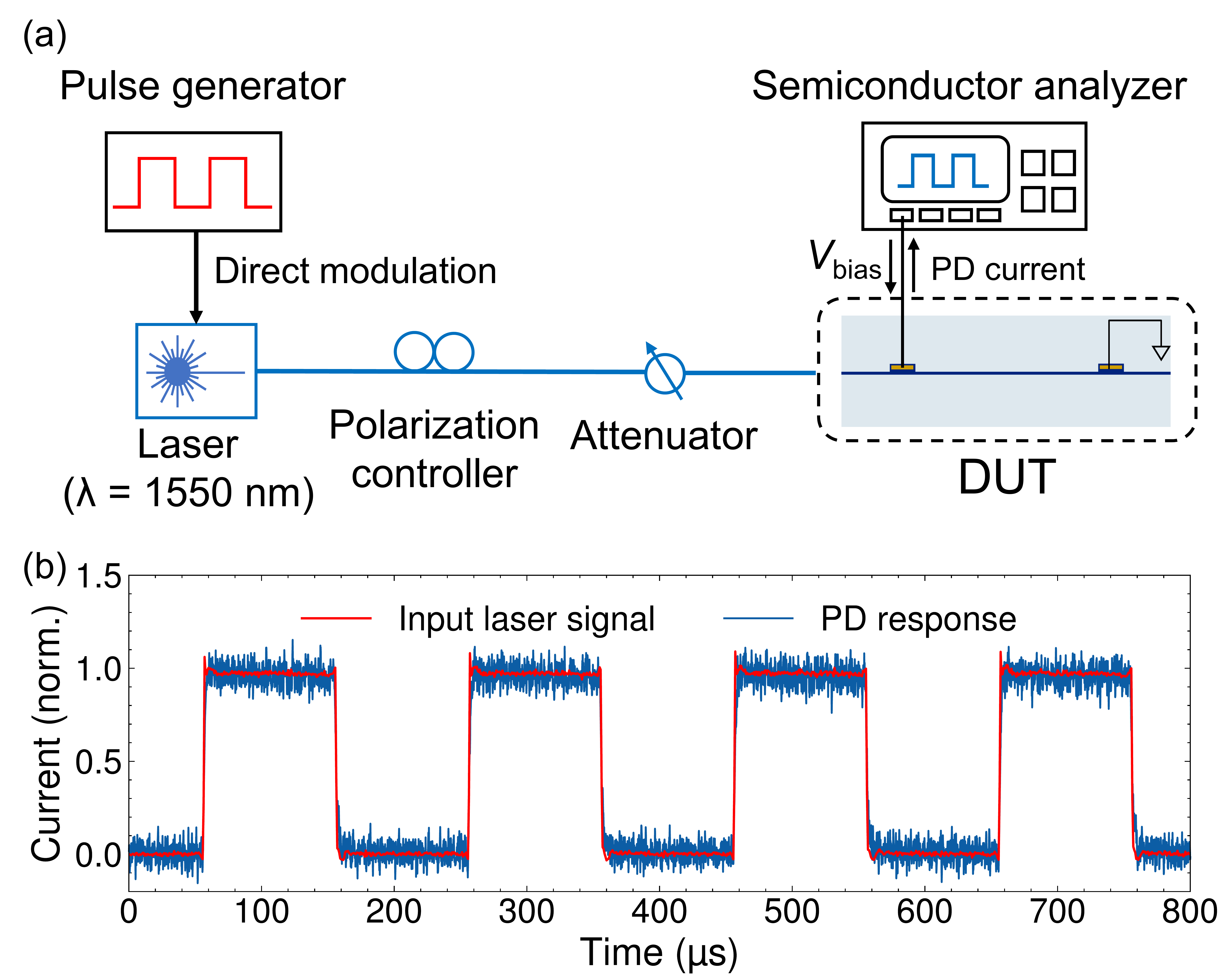}
\caption{(a) Measurement setup for the dynamic photoresponse characterization of a Si WG PD. (b) Time response of a Si WG PD under a bias voltage of 2 V when the input light power to the PD is 4.3 dBm.}
\label{fig:fig3}
\end{figure}
\begin{figure}[ht!]
\centering
\includegraphics[width=70mm]{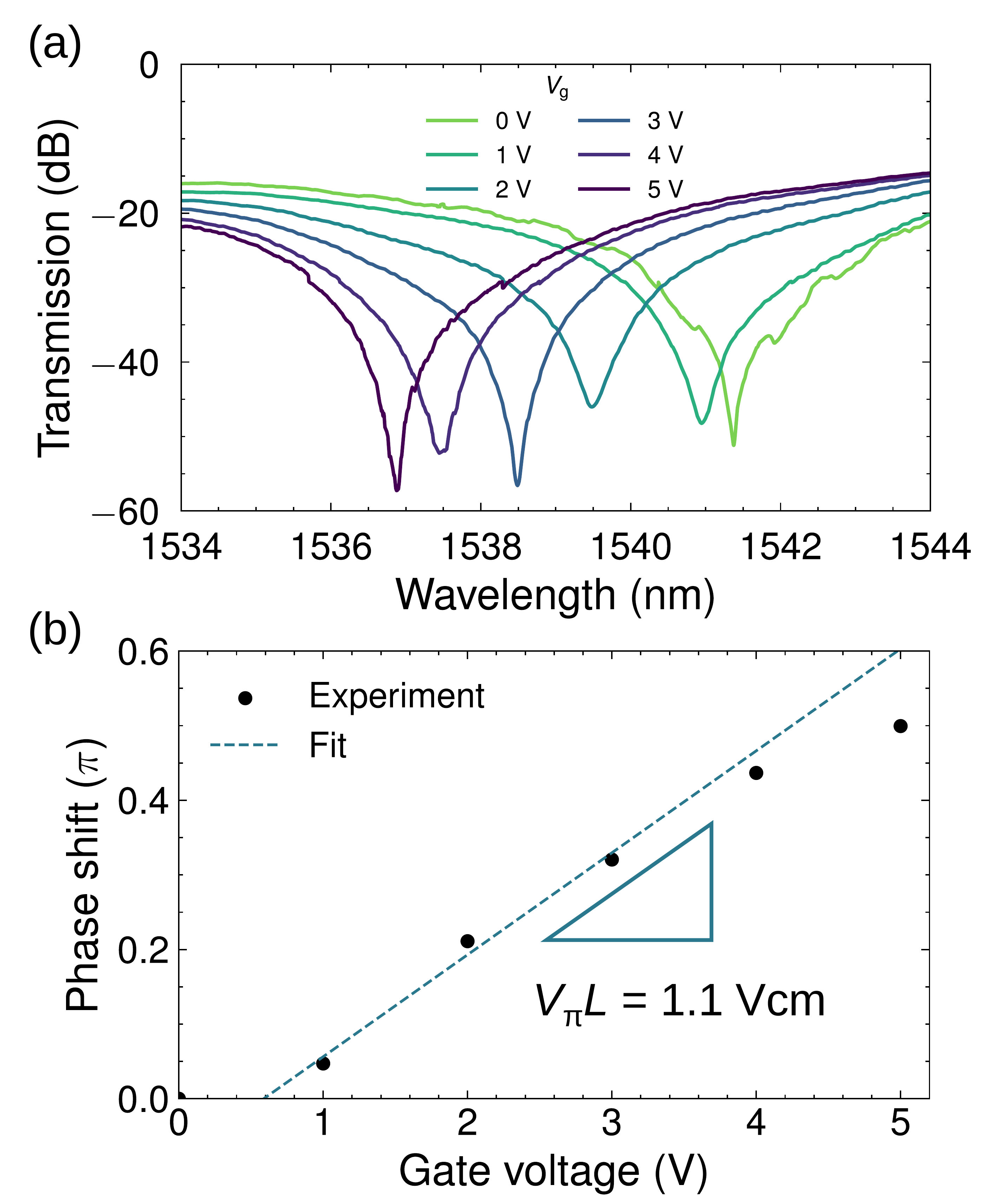}
\caption{(a) Measured transmission spectra of an InGaAsP/Si hybrid MOS optical phase shifter at different gate voltages ($V_\mathrm{g}$). (b) Phase shift as a function of $V_\mathrm{g}$.}
\label{fig:fig4}
\end{figure}
To characterize the InGaAsP/Si hybrid MOS optical phase shifter, we measured the transmission spectra of a 1$\times$1 asymmetric MZI with 1510-µm-long InGaAsP/Si hybrid MOS optical phase shifters at different gate voltages ($V_\mathrm{g}$). As shown in Fig. \ref{fig:fig4}(a), a clear wavelength shift of the resonance peak was observed owing to the electron accumulation at the InGaAsP MOS interface. From this result, the phase shift value at each gate voltage was plotted by calculating the ratio of the wavelength peak shift to a free spectral range (FSR) of 18 nm, as shown in Fig. \ref{fig:fig4}(b). From the slope in the $V_\mathrm{g}$ range from 1 V to 3 V, the modulation efficiency $V_\pi L$ was calculated to be 1.1 Vcm. The discrepancy of the phase shift when $V_\mathrm{g}$ $\geq$ 4 V from the fitted line can be attributed to the large interfacial density at the InGaAsP MOS interface \cite{Han2017}. Considering the thicknesses of the embedded SiO$_2$ and Al$_2$O$_3$, the equivalent oxide thickness (EOT) of the measured MOS capacitor for the optical phase shifter was estimated to be about 18 nm; therefore, the modulation efficiency can be further improved by EOT scaling by thinning the embedded SiO$_2$ \cite{Ohno2020}. 
\begin{figure}[h]
\centering
\includegraphics[width=85mm]{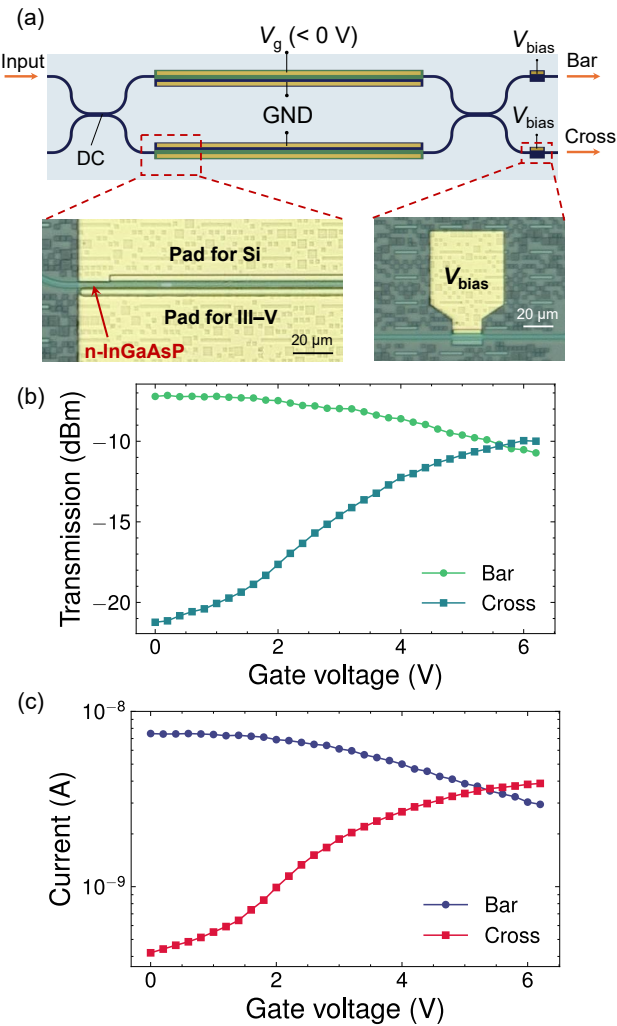}
\caption{(a) Schematic and microscopy images of 2$\times$2 MZI with InGaAsP/Si hybrid MOS optical phase shifters and Si WG PDs. (b) Measured transmission from bar and cross ports as a function of $V_\mathrm{g}$. (c) Measured current at bar and cross WGs as a function of $V_\mathrm{g}$.}
\label{fig:fig5}
\end{figure}
We monolithically integrated the aforementioned Si WG PDs and InGaAsP/Si MOS optical phase shifters in a 2$\times$2 MZI as shown in Fig. \ref{fig:fig5}(a). To monitor the optical power in the two output WGs of the MZI, Si slab regions for metal contacts were formed for both bar and cross WGs. Here, the length of the phase shifters was 1490 µm and the length of the Si slab regions for metal contacts was 20 µm. As shown in Fig. \ref{fig:fig5}(a), CW light at a wavelength of 1550 nm was injected to the upper input WG of the MZI through a GC, and the $V_\mathrm{g}$ applied to the InGaAs/Si hybrid MOS optical phase shifter was then swept to tune the optical power in the bar and cross WGs. Fig. \ref{fig:fig5}(b) shows the measured transmission from bar and cross ports under different $V_\mathrm{g}$ values. Although the bar state was observed when $V_\mathrm{g}$ = 0 V due to the initial phase error by fabrication, the transmission from the cross port increased with $V_\mathrm{g}$. To monitor the optical power in bar and cross WGs under each $V_\mathrm{g}$, bias voltages ($V_\mathrm{bias}$) were applied to the Si slabs for both WGs, as shown in Fig. \ref{fig:fig5}(a). Here, the GND terminals of the MOS optical phase shifter work as the GND terminals for PDs as well, enabling a simple optical power monitoring. Note that we utilized a directional coupler (DC) consisting of two Si rib WGs that were not electrically isolated; therefore, only the difference in optical power between bar and cross WGs was measured as photocurrent. Here, the WG length from the Si slab region to the DC was about 414 µm. Fig. \ref{fig:fig5}(c) shows the measured current at a bias voltage of -2 V as a function of $V_\mathrm{g}$. Note that a negative bias voltage was applied here to ensure that the voltage difference between $V_\mathrm{g}$ and $V_\mathrm{bias}$ is not large, preventing the increase in the gate leakage current of the MOS capacitor. As can be seen from Fig. 5(c), the measured current clearly followed the transmission in each WG, successfully demonstrating the optical power monitoring in the proposed simple configuration.

\textbf{Conclusion.} 
In this work, we proposed a simple optical power monitoring scheme for a 2$\times$2 MZI with InGaAsP/Si hybrid MOS optical phase shifters. We utilized a low-doped p-Si rib WG PD as a transparent in-line OPM with a time response on the order of 1 µs, which is sufficiently fast for the optical power monitoring application. Based on this Si WG PD, we demonstrated that the optical power in the output WGs of a 2$\times$2 MZI based on InGaAsP/Si hybrid MOS optical phase shifters can be simply monitored by applying bias voltage to the Si slabs formed at the output WGs of the MZI. The proposed optical power monitoring scheme enables the simple monitoring of the power splitting ratio of a 2$\times$2 MZI without excess optical insertion loss and without employing additional complex topology into the MZI.

\begin{backmatter}
\bmsection{Funding} Japan Society for the Promotion of Science (JP23H00172, JP24KJ0823); New Energy and Industrial Technology Development Organization (JPNP16007); JST-Mirai Program (JPMJMI20A1); Ministry of Education, Culture, Sports, Science and Technology (JPMXP1224UT1028); Core Research for Evolutional Science and Technology (JPMJCR2004).

\bmsection{Acknowledgments} The device fabrication was partly conducted in Takeda Cleanroom with help of Nanofabrication Platform Center of School of Engineering, the University of Tokyo, Japan. The authors thank Eisaku Kato, Takuo Tanemura, and Yoshiaki Nakano for support in device fabrication.

\bmsection{Disclosures} The authors declare no conflicts of interest.

\bmsection{Data Availability Statement} Data underlying the results presented in this paper may be obtained from the authors upon reasonable request.

\end{backmatter}

\bibliography{OL_Akazawa}

\bibliographyfullrefs{OL_Akazawa}

\end{document}